\documentclass[prd, onecolumn, nofootinbib,notitlepage,floatfix]{revtex4-1}

\usepackage{amsmath}
\usepackage{graphicx}
\usepackage{dcolumn}
\usepackage{bm}
\usepackage{epsfig}
\usepackage{amssymb,latexsym,mathrsfs}
\usepackage{graphicx}
\usepackage{color}
\usepackage{hyperref}

\hypersetup{
    colorlinks=true,
    linkcolor=red,
    citecolor=blue,
} 

\newcommand{\be}{\begin{equation}}
\newcommand{\ee}{\end{equation}}
\newcommand{\bs}{\begin{split}} 
\newcommand{\bea}{\begin{eqnarray}}
\newcommand{\eea}{\end{eqnarray}}

\newcommand{\mpl}{M^2_{\rm  pl}} 
\newcommand{\lamt}{\lambda^3}
\newcommand{\hds}{h} 
\newcommand{\nmax}{n_{\rm max}}
\newcommand{\nmin}{n_{\rm min}}
\newcommand{\pmax}{p_{\rm max}}
\newcommand{\pmin}{p_{\rm min}}

\begin{document}

\title{An Expansion of Well Tempered Gravity} 

\author{Eric V.~Linder$^{1,2}$, Stephen Appleby$^{3}$} 
\affiliation{
$^1$Berkeley Center for Cosmological Physics \& Berkeley Lab, 
University of California, Berkeley, CA 94720, USA\\  
$^2$Energetic Cosmos Laboratory, Nazarbayev University, 
Nur-Sultan, 010000, Kazakhstan \\ 
$^3$Asia Pacific Center for Theoretical Physics, Pohang, 37673, Korea }

\begin{abstract} 
When faced with two nigh intractable problems in cosmology -- how to 
remove the original cosmological constant problem and how to 
parametrize modified gravity to explain current cosmic 
acceleration -- we can make progress by counterposing them. The well tempered solution to the cosmological constant through degenerate 
scalar field dynamics also relates disparate Horndeski gravity 
terms, making them contrapuntal. We derive  
the connection between the kinetic term $K$ and braiding term $G_3$ for 
shift symmetric theories (including the running Planck mass $G_4$), extending previous work on monomial or binomial 
dependence to polynomials of arbitrary finite degree. We also exhibit 
an example for an infinite series expansion. This contrapuntal condition 
greatly reduces the number of parameters needed to test modified 
gravity against cosmological observations, for these ``golden'' theories of gravity. 
\end{abstract}

\date{\today} 


{
\let\clearpage\relax
\maketitle
}

\section{Introduction}

Current cosmological acceleration is an overwhelming characteristic of 
our universe, driving the expansion rate and shutting down the growth of 
large scale structure. Yet in seeking its origin, physics explanations 
almost invariably sweep under the rug the elephant in the room: the 
original cosmological constant problem that a much higher energy scale 
vacuum energy should have dominated the history of the universe, calling 
into question our understanding of how gravity reacts to vacuum energy 
\cite{Weinberg:1988cp,Carroll:2000fy,Martin:2012bt,Nobbenhuis:2004wn,Padilla:2015aaa}. Exploring a low energy cosmic acceleration by 
traversing a physics terrain where there is an elephant under the rug 
is an uncomfortable position. 

Well tempering \cite{us2018,nott2018,us2020a,us2020b} aims to solve the 
original cosmological constant problem by employing a dynamical scalar field 
with certain degeneracy conditions in the equations of motion. To do so 
requires modifications of general relativity, which is also one of the 
favored approaches to explaining current cosmic acceleration. However, 
modified gravity is quite difficult to test against cosmological observations 
in a general manner, without a large number of model dependent assumptions 
and multiple free functions. By contraposing the two problems, well tempering 
shows a path toward resolving them  both. 

Modified gravity often works within Horndeski theory, the most general 
scalar-tensor theory giving second order equations of motion. 
While generally this involves four free functions, we take 
the simplest approach to setting the speed of gravitational waves 
to be equal to the speed of light so that $G_5=0$ and $G_4=G_4(\phi)$. 
Then Horndeski 
gravity involves three free functions $K(\phi,X)$, $G_3(\phi,X)$, $G_4(\phi)$  
of two variables: the scalar field $\phi$ and its canonical kinetic form 
$X=-g^{\mu\nu}\partial_\mu\phi\partial_\nu\phi/2=\dot\phi^2/2$ for a 
homogeneous scalar field in a Robertson-Walker spacetime. 

To test against cosmological data, one could adopt the effective field 
theory approach \cite{gubitosi,eft1,glpv,bellsaw,eft2} that reduces three functions of two variables to 
four functions of time, at the linear perturbation level. This has the 
drawback that one loses information from the nonlinear regime (including 
details of screening to satisfy solar system tests), where 
cosmological data can give great insight. If we wish to keep the full 
leverage, one 
generally has to assume a specific 
gravity theory, a specific functional dependence within that theory, and 
specific parameters within those functions. For example, one does not 
``test gravity'' but says the data is or is not consistent with, say, 
``the class of $f(R)$ gravity with a specific function $f$ and specific 
ranges of parameters appearing within that function $f$.'' The conclusions tend to be 
highly model dependent. To put the general case of three functions of 
two variables in perspective, figuring out how to modify gravity yet 
obtain consistency with cosmological observations is like trying to 
figure out how to kick a football to score a goal when 1) the pull of 
gravity depends on the ball's location and velocity, 2) the wind speed 
varies with the ball's location and velocity, and 3) the ball's mass 
depends on its location (e.g.\ $K(\phi,X)$, $G_3(\phi,X)$, $G_4(\phi)$ 
respectively). 

In \cite{us2020a} we demonstrated that well tempering could not only address  
the original cosmological constant problem, but reduce the modified 
gravity parameter space from multiple functions to a small number  
($\sim$ four) of parameters. These give well defined, valuable theories 
that are predictive across the full range of cosmological observations, 
including on nonlinear scales, and could be regarded as the ``golden'' 
theories to use in cosmology, without climbing over the elephant under 
the rug. 

The solutions in \cite{us2020a} were predominantly within shift symmetric 
theories, which have important protection against quantum corrections \cite{Finelli:2018upr,Heisenberg:2018vsk},
and gave particular solutions to the well tempered degeneracy equation by 
assuming various Ans{\"a}tze for either $K(X)$ or $G_3(X)$, with the 
other then determined by the equations. The degeneracy equation is a 
nonlinear differential equation and so other solutions can exist as 
well. Here we carry out a systematic analysis of all solutions having 
certain conditions discussed below. 

In Section~\ref{sec:series} we restate the key degeneracy equation 
giving well tempering and introduce a series expansion method. The 
remarkably simple solutions for a finite series are presented in  
Sec.~\ref{sec:finite} and we discuss some special cases as  well, and how previous solutions 
are unified by our general expression. 
We present an 
example of an infinite series solution in Sec.~\ref{sec:infinite}. 
We summarize and conclude in Sec.~\ref{sec:concl}.

\section{Series Expansion Approach} \label{sec:series} 

Well tempering works by reducing the equations of motion involving 
$\ddot a$ and $\ddot\phi$, i.e.\ the expansion and scalar field 
evolutions, to have degenerate solutions for de Sitter spacetime, 
referred to as ``on shell''. This empowers the scalar field dynamics 
to cancel the cosmological constant both on and off shell, i.e.\ for 
the full cosmological history. See \cite{us2018} for details.  

The equation guaranteeing the degeneracy is 
\bea 
& &(M-2g)\left\{3\hds\dot\phi K_X+18\hds^2 g-6\hds^2 M+\lamt\right\}\notag\\ 
& &\qquad= (K_X+2XK_{XX}+6\hds\dot\phi g_X) \left[-\hds\dot\phi(M-6g)+2XK_X\right]\,, \label{eq:degen} 
\eea 
where we have imposed shift symmetry so that $G_4=(\mpl+M\phi)/2$, 
and $G_3=G_3(X)$, $K(\phi,X)=F(X)-\lamt\phi$. For notational 
convenience we write $g=XG_{3X}$, where a subscript $X$ denotes a 
derivative with respect to $X$, and $h\equiv H_{\rm dS}$ is the de Sitter value 
of the expansion rate. 

This nonlinear differential equation relates the Horndeski terms 
$K$, $G_3$, and $G_4$. A variety of solutions were presented and 
analyzed in \cite{us2020a}. Here we look for more general 
solutions using a series expansion method like that introduced and 
used to good effect in \cite{nott2018}. Note that in their work 
they assumed $K=X-V(\phi)$, which  
turns the nonlinear equation into a simpler one (but 
their nonassumption of shift symmetry makes it a partial differential equation). 
Here we impose shift symmetry, justified for its protection against 
quantum loop corrections, but allow general $K(X)$ with the permitted $\lamt\phi$ tadpole term. 

By resummation and judicious insight, a series expansion can 
provide the general functional dependence $K(X)$ or $G_3(X)$. 
This method can deliver an algorithm for generating new 
solutions, and some general principles. 

Since the quantities entering the degeneracy equation are $K_X$ and $g$, 
we will carry out series expansion of these; they can be rewritten in 
terms of $K$ and $G_3(X)$ as desired. Thus, 
\bea 
g&=&\sum a_n X^{n/2}\\ 
K_X&=&\sum b_p X^{p/2}\,. 
\eea 
The sums go from $n_{\rm min}$ to $n_{\rm max}$ and $p_{\rm min}$ to 
$p_{\rm max}$ respectively, and we will see there are relations between 
these values. The degeneracy equation becomes 
\bea 
0&=& 6h^2M\sum a_n \left(5+n-\frac{\lamt}{3h^2M}\right)\,X^{n/2}
-36h^2\left(\sum a_n X^{n/2}\right)\,\left(\sum a_n(1+n)X^{n/2}\right)\notag\\ 
&+&Mh\sqrt{2X}\sum b_p (4+p)X^{p/2} 
-2X\left(\sum b_p X^{p/2}\right)\,\left(\sum b_p(1+p)X^{p/2}\right)\notag\\ 
&-&6h\sqrt{2X}\left[\left(\sum a_n X^{n/2}\right)\,\left(\sum b_p(1+p)X^{p/2}\right)+\left(\sum a_n(1+n) X^{n/2}\right)\,\left(\sum b_p X^{p/2}\right)\right]\notag\\ 
&-&6h^2M^2+M\lamt\,. \label{eq:degdbl} 
\eea

\section{Finite Series Solution} \label{sec:finite} 

One solves Eq.~\eqref{eq:degdbl} by equating terms order 
by order in $X$ to obtain relations between 
$a_n$ and $b_p$, and then summing the series to derive 
functional relations between $K$ and $G_3$ (or $g$), 
with or without $M$ from the $G_4$ term. We begin by 
considering a finite series.

\subsection{General Results} \label{sec:genfin} 

The results are 
\bea 
K_X &=&-3h\sqrt{2}X^{-1/2}g +h\sqrt{2}X^{-1/2}\left(M-\frac{\lamt}{6h^2}\right) \qquad {\rm[Branch\ A]} \label{eq:kxa}\\  
K_X &=&-3h\sqrt{2}X^{-1/2}\left(g-a_0\right)-\frac{3h\sqrt{2}}{2}\,X^{-1/2}\int dX\,\frac{g-a_0}{X} +b_{-1}X^{-1/2} \qquad {\rm[Branch\ B]} \,.  \label{eq:kxb}
\eea 
Appendix~\ref{sec:apxfinite} shows the steps in the 
derivation but one can verify the solutions by direct 
substitution into Eq.~\eqref{eq:degen}. 

The coefficient $b_{-1}$ entering as $K_X^{[-1]}=b_{-1}X^{-1/2}$ plays a special role 
(since $K_X^{[-1]}+2XK_{XX}^{[-1]}=0$ in the degeneracy equation) and helps distinguish the two branches. For Branch A, 
\be   
b_{-1}=h\sqrt{2}M-\frac{\lamt}{3h\sqrt{2}}-3h\sqrt{2}\,a_0 \qquad [{\rm Branch\ A}]\,, 
\label{eq:b1a} 
\ee  
i.e.\ $b_{-1}$ is specifically connected to $a_0$, 
the constant term in $g$ or the contribution 
$G_3^{[0]}=a_0\ln  X$ (recall from \cite{us2020a} that 
constant $g$ is a hallmark of Brans-Dicke type 
scalar-tensor theories as well as $f(R)$ gravity and 
No Slip Gravity). 
The quantity $b_{-1}$ is also 
connected to the $G_4$ mass $M$ and the tadpole 
scale $\lambda$ for Branch A, so all the functions are 
woven together. 

Branch B has different contrapuntal conditions between the functions, 
with 
\be 
b_{-1}=\{{-h\sqrt{2}M,\ \rm  arbitrary}\}, \quad a_0=\frac{M}{2}, \quad \lamt=3h^2M \qquad [{\rm Branch\ B}]\,. 
\label{eq:b1b} 
\ee 
Here the connections are more concrete between $G_3$, 
$G_4$, and $K$, i.e.\ $a_0$, $M$, and $\lambda$, but 
$b_{-1}$ also has the possibility of being arbitrary, 
as we  discuss in the next section. 

Equations~\eqref{eq:kxa} and \eqref{eq:kxb} are general solutions. They cover many of the  
particular solutions given in \cite{us2020a} (hereafter called HV).  For example, HV Eq.~3.36 is a Branch A solution; Eqs.~3.34 and 3.38 
are Branch B solutions. We have essentially 
succeeded in deriving a unified solution for all 
the individual solutions obtained 
in HV. 

Some particular solutions need or merit special treatment 
and we deal with these in the next subsections.

\subsection{Special Case: Truncation Below $b_{-1}$}  \label{sec:trunc} 

Under the standard 
solutions of the previous section, for some $a_n$, even 
$a_{\nmin}$, there must be a $b_{n-1}$, 
as shown in Eq.~\eqref{eq:bpup}. 
However, we see that the $a^2_{\nmin}$ contribution (the second term in  
Eq.~\ref{eq:degdbl}) vanishes when $n=-1$ and so this term is an exception: 
if we set $\nmin=-1$ we are free to set $b_{-2}=0$ 
(and all $b_{p<-2}$ are zero likewise). 
This may be considered attractive in that it 
means that $K$ does not have any negative powers of  
$X$, so when the field rolls slowly there is no 
blow up of the kinetic term. 
Thus we consider this special case $\nmin=-1$, i.e.\ $a_{n<-1}=0$. 

With $b_{-2}=0$, 
for the next higher order equation, which involves $X^{-1/2}$ terms, $b_{-1}$ 
actually cancels out of the degeneracy equation. We continue going 
through the intermediate powers and find that for Branch A, $b_{-1}$ takes the 
form in Eq.~\eqref{eq:b1a}, {\it unless\/} $a_{n>0}=0$ and $a_0=M/2$, in 
which case $b_{-1}$ remains arbitrary. The value of $a_{-1}$ does not  affect 
$b_{-1}$ under these conditions. For Branch B, Eq.~\eqref{eq:b1a} reduces 
to $b_{-1}=-h\sqrt{2}M$; however, if the above conditions hold and we also 
require $a_{-1}=0$, then $b_{-1}$ is arbitrary. 
This explains the ``$b_{-1}$ arbitrary'' possibility mentioned 
in the previous subsection. 

Cases with arbitrary $b_{-1}$ 
can be seen in HV Eqs.~3.31, 3.35, and 3.39, all Branch B solutions, 
and Eq.~3.37, a Branch A solution but where the arbitrary $b_{-1}$ arises  
from an arbitrary term in $a_0$, i.e.\ it still follows Eq.~\eqref{eq:b1a}.

\subsection{Special Case: $\nmin=0$, $\nmax=1$} \label{sec:01} 

One other special case of note is truncation where 
$\nmin=0$ and in addition $\nmax=1$, i.e.\ $g=sX^{1/2}+a_0$. 
This gives only 
three equations, so in addition to determining 
$b_{-1}$ and  $b_0$ they must fix $a_0$. 
The result shows that despite the tight restrictions this case  
nevertheless follows the general  Eqs.~\eqref{eq:kxa} and \eqref{eq:kxb}. 
That is, Branch A gives  
\bea 
g&=&sX^{1/2}+\frac{M}{3}-\frac{b_{-1}}{3h\sqrt{2}}-\frac{\lamt}{18h^2}\\ 
K_X&=&-3h\sqrt{2}s+b_{-1}X^{-1/2}\,, 
\eea 
with $b_{-1}$ arbitrary and $\lamt\ne 3h^2M$, following from Eq.~\eqref{eq:kxa}, 
and Branch B, i.e.\ Eq.~\eqref{eq:kxb}, yields  
\bea 
g&=&sX^{1/2}+\frac{M}{3}+\frac{\lamt}{18h^2}\\  
K_X&=&-6h\sqrt{2}s-\frac{\lamt\sqrt{2}}{3h}X^{-1/2}\,. 
\eea

We can show this as follows. 
The 
$X^1$ equation gives $b_0=\{-3,-6\}h\sqrt{2}\,a_1$,  
as expected from Eq.~\eqref{eq:bpdown}; 
for the choice 
$b_0=-3h\sqrt{2}\,a_1$, i.e.\ following Branch A, 
the $X^{1/2}$ 
equation gives Eq.~\eqref{eq:b1a} for $b_{-1}$, but 
for the choice $b_0=-6h\sqrt{2}\,a_1$ then $b_{-1}$ 
cancels out 
and instead the equation imposes 
\be 
a_0=\frac{M}{3}+\frac{\lamt}{18h^2}\,, 
\ee 
so this is a hidden version of Branch B. The hidden 
aspect arises because of the lack of extra equations 
that would impose the usual additional  consistency conditions such that 
$\lamt=3h^2M$ and hence $a_0=M/2$. Finally, the 
$X^0$ equation determines $b_{-1}$ following 
Eq.~\eqref{eq:b1a}, which would only lead to Eq.~\eqref{eq:b1b} 
upon setting the additional consistency conditions $a_0=M/2$, $\lamt=3h^2M$, which do not 
apply here. Thus we have $g=sX^{1/2}+a_0$ leading to 
HV Eq.~3.37 for Branch A and HV Eq.~3.33 for the 
``hidden'' Branch B.

\subsection{Special Case: $\ln X$ Terms in $K_X$} \label{sec:log} 

A finite expansion in powers will not always work. 
Before we move on to an infinite series in  
Sec.~\ref{sec:infinite}, let us consider the 
special case where $K_X$ contains terms involving 
$\ln X$. This is particularly of note since 
\cite{us2020a} did find solutions  
involving $\ln X$. 
Equations~\eqref{eq:klnx} and 
\eqref{eq:glnx} below provide the 
final solutions. 
To derive them we begin by expanding $g$ 
and studying the form of the resulting degeneracy 
equation, writing  
\bea  
g&=&\sum a_n X^{n/2}\\ 
K_X&=&X^{-1/2}\,B(X)\,, 
\eea  
where $K$ still has a tadpole term  $-\lamt\phi$. 

The degeneracy equation becomes 
\bea 
0&=&XB_X\left(4B+12h\sqrt{2}\sum a_n  X^{n/2}-2h\sqrt{2}M\right)+
h\sqrt{2}B\left[6\sum a_n(1+n) X^{n/2}-3M\right]\\ 
&\qquad&-6h^2M\sum a_n\left(5+n-\frac{\lamt}{3h^2M}\right) 
+36h^2\left(\sum a_n X^{n/2}\right)\left(\sum a_n(1+n)X^{n/2}\right) 
+6h^2M^2-M\lamt\,. \notag
\eea 
A term involving $B\sim X^m\ln X$ has nothing to cancel against  
if $m\ne0$, so we allow only a term like $B\sim \ln X$, besides standard powers. 
Writing 
\be 
B=b_r\,\ln  X+\sum b_p  X^{p/2}\,, 
\ee 
the degeneracy equation for the terms involving $\ln  X$ becomes 
\be 
0=4b_r\ln X\,\left(b_r+\frac{1}{2}\sum b_p pX^{p/2}\right)+h\sqrt{2}b_r\ln  X\, 
\left[6\sum a_n(1+n)X^{n/2}-3M\right]\,. 
\ee 
If we set $b_r=0$, there is no $\ln  X$ term and we return to the 
power series of the previous sections. The solution for general $p\ne 0$, $-1$ 
is 
\be  
a_p=b_p\,\frac{p}{3h\sqrt{2}(1+p)}\,. 
\ee 
However, looking at the terms in the degeneracy 
equation involving only powers and not $\ln X$, i.e.\ $X^{p}$, 
we find the only consistent solution is $b_p=0$ for all $p\ne0$. 
For $p=0$, we find $b_0$ is arbitrary and 
\be 
a_0=\frac{M}{2}-\frac{b_r\sqrt{2}}{3h}\,. 
\ee 
For the $X^0$ order (without $\ln X$), the solution requires either 
$a_0=M/2$ (implying $b_r=0$ and hence reducing to  the pure power 
expansion without $\ln X$ as in the previous sections) or 
$\lamt=3h^2M$. 

Thus the final solution is $B=b_0+b_r\ln X$, or 
\be   
K_X=b_0 X^{-1/2}+b_rX^{-1/2}\ln X\,, \label{eq:klnx} 
\ee 
with a tadpole term $-3h^2M\phi$ in $K$, 
and  
\be 
g=\frac{M}{2}-\frac{b_r\sqrt{2}}{3h}+a_{-1}X^{-1/2}\,, \label{eq:glnx}
\ee 
where $a_{-1}$ is arbitrary. 
This is the unique solution where $K_X$ involves 
a $\ln X$ term, and is equivalent to HV Eq.~3.20. 
When $g=0$, so $b_r=3hM/(2\sqrt{2})$, we reproduce HV Eq.~3.11, and  
when $g=rM$ (as for $f(R)$ and 
No Slip Gravity), we obtain HV Eq.~3.19.

\section{Infinite Series Solution} \label{sec:infinite} 

When the series expansion is infinite then one must use 
the Cauchy product to determine terms at a certain 
order. Since there is no finite $\nmax=N$, we are no longer 
constrained by the lack of a counter term to $a_N^2 X^N$. 
This breaks the relation between $a_n$ and 
$b_{n-1}$. There is very little one can say in general 
for such a situation. However, if one restricts the 
series in some way then some progress can be made. For 
example, consider the case where $b_{n\ne-1}=0$. Then 
the only series product comes from the second term in  
Eq.~\eqref{eq:degdbl}, which we evaluate using the Cauchy product, 
except for separating out where one term has  
$a_{-1}X^{-1/2}$. If we choose $a_n$ as a semi-infinite 
series, with $a_{n<-1}=0$, then we can solve the 
equation. 

Starting with the $X^{-1/2}$ power equation and working 
up, we can obtain all $a_n$ and hence $g$. A particularly 
compact form obtains  
for $b_{-1}=-\lamt\sqrt{2}/(3h)$, in that then the 
(semi)infinite series sums to the solution  
\be 
g=\frac{M}{2}+\frac{c}{2}\,X^{-1/2}\left[1\pm\sqrt{1+\frac{2(3h^2M-\lamt)}{9h^2c}X^{1/2}}\ \right]\,. \label{eq:345}
\ee 
Note that the lower root gives $a_{-1}=0$, and hence 
$g$ involves only nonnegative powers of $X$. 
(For the upper root $c=a_{-1}$.) When 
$\lamt=0$, the 
entire kinetic term $K=2b_{-1}X^{1/2}-\lamt\phi=0$ 
and we have HV Eq.~3.45. When $\lamt>3h^2M$ we have 
a ``speed limit'' on $X$, i.e.\ the scalar field motion 
$\dot\phi$, to keep the function real 
(see for example \cite{eva1,eva2,eva3}).

\section{Conclusions} \label{sec:concl} 

The reduction of the description of modified gravity from three 
functions of two variables to a single function of one variable, or 
a handful of constant parameters, could open up powerful leverage on 
scanning theory space to compare to observational data. Remarkably, 
we have shown this can be done with the bonus of solving the original 
cosmological constant problem -- through well tempering -- and 
protecting from at least some quantum corrections -- through shift 
symmetry. By contraposing two highly challenging problems we solve 
both. 

Equations~\eqref{eq:kxa} and \eqref{eq:kxb} give general solutions 
relating the Lagrangian terms $K$, $G_3$, and $G_4$ for a wide range 
of gravity theories, derived using a power series expansion under 
well tempering. 
We show that they unify the disparate solutions 
found piecemeal in \cite{us2020a}, while going well beyond them, 
extending monomial or binomial cases to arbitrary finite polynomials. 
Logarithmic terms are included as well. Branch A solutions are what 
\cite{us2020a} referred to as the $(\checkmark)$ class that gives 
a scalar field equation that becomes trivial on shell, 
while Branch B solutions are fully well tempering. 

Allowing the series expansion to be infinite gives formal 
solutions but ones difficult to sum to a compact functional form. 
We exhibit one example where this can be done, generalizing a 
case from \cite{us2020a}. For this result, the full gravity theory 
can be described by four constant parameters: $M$, $\lambda$, $h$, 
and $c$, which can readily be sampled for likelihood estimation 
compared to data. 

Such theories that possess highly desirable characteristics for  
fundamental physics -- solving rather than neglecting the cosmological 
constant problem, and protecting against quantum corrections --   
soundness, and robust ability for full comparison with observations 
(including nonlinear scales, in principle) could be regarded as the 
favored ``golden'' gravity theories to work with. While exciting  
work remains to investigate further their detailed properties, they 
represent a significant step away from arbitrary functions toward 
true benchmarks.

\acknowledgments 
SAA is supported by an appointment to the JRG Program at the APCTP through the Science and Technology Promotion Fund and Lottery Fund of the Korean Government, and was also supported by the Korean Local Governments in Gyeongsangbuk-do Province and Pohang City. 
EL is supported in part by the Energetic Cosmos Laboratory and by the U.S. Department of Energy, Office of Science, Office of High Energy Physics under contract no. DE-AC02-05CH11231.

\appendix 

\section{Derivation for Finite Series} \label{sec:apxfinite} 

To begin the analysis of the finite series case 
order by order, let us 
look at the maximum powers for each series. 
For $N\equiv\nmax$ we have one term going as $X^N$. To match this we must have 
$P\equiv\pmax=N-1$, unless $N=0$. That is, the $K_X$ series must cut off 
at one less power than the $g$ series.  We find 
\be 
b_{N-1}=-3h\sqrt{2}\, a_N\left(1+\frac{1\mp1}{2N}\right)=
\left\{-3h\sqrt{2}\, a_N,\,-3h\sqrt{2}\, a_N\frac{N+1}{N}\right\}\,. 
\ee 
Thus there are two branches of solutions, what we call Branch A and Branch B. 
We can then proceed to the next lowest power, $X^{N-1/2}$ and determine $b_{N-2}$ from $a_{N-1}$. 
By continuing this process for $X^n$, for $N/2<n\le N$, we find 
\be 
b_p=\left\{-3h\sqrt{2}\, a_{p+1},\,-3h\sqrt{2}\, a_{p+1}\,\frac{p+2}{p+1}\right\}\qquad  
{\rm for} \qquad 0\le p\le N-1\,. \label{eq:bpdown} 
\ee 
There are $N$ equations, for $X^N$, $X^{N-1/2}$,\dots $X^{(N+1)/2}$, and these 
define the $N$ coefficients $b_{N-1}$, $b_{N-2}$,\dots $b_0$. 
Now let us jump to the most negative power of $X$, 
i.e.\ $L\equiv\nmin<0$ and evaluate the $b_p$ with $p<0$ working 
upward. One obtains basically the same expression as Eq.~(\ref{eq:bpdown}), and 
again $\pmin=\nmin-1$: 
\be 
b_p=\left\{-3h\sqrt{2}\, a_{p+1},\,-3h\sqrt{2}\, a_{p+1}\,\frac{p+2}{p+1}\right\}\qquad  
{\rm for} \qquad L-1\le p<-1\,. \label{eq:bpup} 
\ee 
Again there are two branches, and we must choose the same branch for the  
negative powers as the positive powers.  
These determine the $L$ coefficients $b_{L-1}$,  $b_{L}$,\dots $b_{-2}$. 

There are several equations for the powers $X^n$ with $-L/2\le n\le N/2$ and only a 
single parameter $b_{-1}$ left to determine. All these equations must give a 
consistent solution for $b_{-1}$. 
Note that more and more terms from Eq.~(\ref{eq:degdbl}) 
enter into these equations, but a consistent solution occurs 
for each branch. For branch A,  
we need 
\be   
b_{-1}=h\sqrt{2}M-\frac{\lamt}{3h\sqrt{2}}-3h\sqrt{2}\,a_0 \qquad [{\rm Branch\ A}]\,, 
\label{eq:apxb1a} 
\ee  
and for branch B, the requirements are 
\be 
b_{-1}=\{{-h\sqrt{2}M,\ \rm  arbitrary}\}, \quad a_0=\frac{M}{2}, \quad \lamt=3h^2M \qquad [{\rm Branch\ B}]\,.  
\label{eq:apxb1b} 
\ee 

To summarize, all $b_{p}$ are determined by Eq.~\eqref{eq:bpdown} 
(or equivalently Eq.~\ref{eq:bpup}), except for $b_{-1}$ -- special since 
for this order $K_X+2XK_{XX}=0$ -- which is given by Eq.~\eqref{eq:apxb1a} or 
\eqref{eq:apxb1b} for the respective branch.  

We can now try to sum up the series to obtain a functional relation. 
For Branch A, 
\bea
K_X&=&\sum b_p X^{p/2}=-3h\sqrt{2}\sum_{p\ne-1} a_{p+1}X^{p/2}+b_{-1}X^{-1/2}=-3h\sqrt{2}X^{-1/2}\sum_{p\ne-1} a_{p+1}X^{(p+1)/2} +b_{-1}X^{-1/2}\notag\\ 
&=&-3h\sqrt{2}X^{-1/2}\sum_{n\ne0} a_{n}X^{n/2} +b_{-1}X^{-1/2}\notag\\ 
&=&-3h\sqrt{2}X^{-1/2}g +h\sqrt{2}X^{-1/2}\left(M-\frac{\lamt}{6h^2}\right) \qquad {\rm[Branch\ A]} \,. 
\label{eq:apxkxa}
\eea 
For Branch B,   
\bea 
K_X&=&-3h\sqrt{2}\sum_{p\ne-1} \frac{p+2}{p+1}\,a_{p+1}X^{p/2}+b_{-1}X^{-1/2}=-3h\sqrt{2}X^{-1/2}\sum_{p\ne-1} \frac{p+2}{p+1}\,a_{p+1}X^{(p+1)/2} +b_{-1}X^{-1/2}\notag\\  
&=&-3h\sqrt{2}X^{-1/2}\sum_{n\ne0}  \left(1+\frac{1}{n}\right)\,a_{n}X^{n/2} +b_{-1}X^{-1/2}\notag\\ 
&=&-3h\sqrt{2}X^{-1/2}\left(g-a_0\right)-\frac{3h\sqrt{2}}{2}\,X^{-1/2}\int dX\,\frac{g-a_0}{X} +b_{-1}X^{-1/2} \qquad {\rm[Branch\ B]} \,.  \label{eq:apxkxb}
\eea 
Recall that for Branch B, $\lamt=3h^2M$ and the constant part of $g$ is simply $a_0=M/2$. 
Thus we have derived our  general solutions Eqs.~\eqref{eq:kxa} and \eqref{eq:kxb}. 

Substituting these back into the degeneracy Eq.~\eqref{eq:degen}, we find that Branch A 
gives a zero for the first factor on the right hand side, $K_X+2XK_{XX}+6h\dot\phi g_X$. 
This indicates the coefficient of $\ddot\phi$ in the scalar field equation vanishes 
on shell. This is what we called $(\checkmark)$ models in \cite{us2020a}. Branch B does not zero 
out coefficients of $\ddot\phi$ and gets a full 
$\checkmark$. 

Note that if desired we can set $M=0$, removing the coupling to the Ricci scalar and making the $G_4$ term standard. For Branch B this will also make $\lamt=0$ and $a_0=0$.



\begin{thebibliography}{99}

\bibitem{Weinberg:1988cp}
S.~Weinberg,
The Cosmological Constant Problem,
Rev. Mod. Phys. \textbf{61}, 1  (1989)

\bibitem{Carroll:2000fy}
S.~M.~Carroll,
The Cosmological constant, 
Living Rev. Rel. \textbf{4}, 1 (2001)
[arXiv:astro-ph/0004075] 

\bibitem{Martin:2012bt}
J.~Martin,
Everything You Always Wanted To Know About The Cosmological Constant Problem (But Were Afraid To Ask), 
Comptes Rendus Physique 13, 566 (2012)
[arXiv:1205.3365] 

\bibitem{Nobbenhuis:2004wn}
S.~Nobbenhuis,
Categorizing different approaches to the cosmological constant problem,
Found. Phys. 36, 613 (2006)
[arXiv:gr-qc/0411093]

\bibitem{Padilla:2015aaa}
A.~Padilla,
Lectures on the Cosmological Constant Problem,
arXiv:1502.05296 

\bibitem{us2018} 
S. Appleby, E.V. Linder, 
The Well-Tempered Cosmological Constant, 
JCAP 1807, 034 (2018) [arXiv:1805.00470] 

\bibitem{nott2018} 
W.T. Emond, C. Li, P.M. Saffin, S-Y. Zhou, 
Well-Tempered Cosmology, 
JCAP 1905,  038 (2020) [arXiv:1812.05480] 

\bibitem{us2020a} 
S. Appleby, E.V. Linder, 
The Well-Tempered Cosmological Constant: The Horndeski Variations, 
JCAP in press [arXiv:2009.01720] 

\bibitem{us2020b} 
S. Appleby, E.V. Linder, 
The Well-Tempered Cosmological Constant: Fugue in B$^\flat$, 
JCAP in press [arXiv:2009.01723] 

\bibitem{gubitosi} 
G. Gubitosi, F. Piazza, and F. Vernizzi, The effective field theory of 
dark energy, JCAP 1302, 032 (2013) [arXiv:1210.0201] 

\bibitem{eft1} 
J.K. Bloomfield, E.E. Flanagan, M. Park, S. Watson, Dark energy or 
modified gravity? An effective field theory approach, JCAP 1308, 010 
(2013) [arXiv:1211.7054] 

\bibitem{glpv} 
J. Gleyzes, D. Langlois, F. Piazza, F. Vernizzi, Essential building 
blocks of dark energy, JCAP 1308, 025 (2013) [arXiv:1304.4840] 

\bibitem{bellsaw} 
E. Bellini, I. Sawicki, Maximal freedom at minimum cost: linear large-scale 
structure in general modifications of gravity, JCAP 1407, 050 (2014) 
[arXiv:1404.3713] 

\bibitem{eft2} 
E.V. Linder, G. Seng{\"o}r, S. Watson, Is the Effective Field Theory of 
Dark Energy Effective?, JCAP 1605, 053 (2016) [arXiv:1512.06180] 

\bibitem{Finelli:2018upr}
B.~Finelli, G.~Goon, E.~Pajer and L.~Santoni,
The Effective Theory of Shift-Symmetric Cosmologies,
JCAP 1805, 060 (2018)
[arXiv:1802.01580] 

\bibitem{Heisenberg:2018vsk}
L.~Heisenberg,
A systematic approach to generalisations of General Relativity and their cosmological implications,
Phys. Rept. 796, 1 (2019)
[arXiv:1807.01725] 

\bibitem{eva1} 
E. Silverstein, D. Tong, Scalar speed limits and cosmology: Acceleration from D-cceleration, Phys. Rev. D 70, 103505 (2004) [arXiv:hep-th/0310221] 

\bibitem{eva2} 
M. Alishahiha, E. Silverstein, D. Tong, DBI in the sky, Phys. Rev. D 70, 123505 (2004) [arXiv:hep-th/0404084] 

\bibitem{eva3} 
D. Mathis, A. Mousatov, G. Panagopoulos, E. Silverstein, 
A new branch of inflationary speed limits, arXiv:2010.00113 


\end{thebibliography}
\end{document}